# SOLUTIONS OF THE SCHRODINGER EQUATION FOR MODIFIED MOBIUS SQUARE POTENTIAL USING TWO APPROXIMATIONS SCHEME.

C. M. Ekpo[1], J. E. Osang[1,2,3,4] and E. B. Ettah[1]

[1]Department of Physics, Cross River University of Technology, Calabar

[2]Department of Physics, River State University Port Harcourt, Nigeria.

[3]Cross River State Emergency Management Agency (SEMA)

[4]Department of Physics, Arthur Javis University, Cross River State, Nigeria

## ABSTRACT

In this paper, the Schrodinger equation for s-wave ($l = 0$) and $l \neq 0$ with the Modified Mobuis Square potential is investigated respectively. The eigenfunctions as well as energy eigenvalues are obtained in an exact analytical manner via the Nikiforov Uvarov method using two approximations scheme. Some special cases of this potentials are also studied.

## 1. INTRODUCTION

It is well known that Bound state solutions of Schrodinger equation play an integral role in Quantum Mechanics. Bound state solutions describe the system of a particle subjected to a potential, with the tendency to remain in a fixed region of space [1]. Quantum mechanical Wavefunctions and their corresponding eigenvalues give significant information in describing various quantum systems [2-19]. Ekpo et al. [20] obtained the approximate analytical solutions of the radial Schrodinger equation for the New Generalized Morse-Like Potential in arbitrary dimensions by using the Nikiforov Uvarov Method. Energy eigenvalues and corresponding eigenfunction are obtain analytically. The approximate analytical solution of the Schrodinger equation with modified Mobius square potential (MMSP) has not been studied by many researchers

$$V(r) = -V_0 \left(\frac{A+Be^{-2\alpha r}}{1-e^{-2\alpha r}}\right)^2 \tag{1a}$$

where $V_0, A, B, and\ \alpha$ are the depth of the potential, the range of the potential, the length of the molecular bond, and an adjustable screening parameter, respectively.

Equation can be rewritten as;

$$V(r) = -V_0 \left(\frac{A^2+2ABe^{-2\alpha r}+B^2 e^{-4\alpha r}}{(1-e^{-2\alpha r})^2}\right) \tag{1b}$$

Okorie et al[21] noted that the MMSP is a general case of the Hulthen and Morse potentials [22, 23]. Okorie *et al* [21] solved the Schrödinger equation with the modified Mobius square potential model using the modified factorization method. This was achieved within the framework of the Greene–Aldrich approximation for the centrifugal term and using a suitable transformation scheme, they obtained the energy eigenvalues equation and the corresponding eigenfunction in terms of the hypergeometric function. Using the resulting eigenvalues equation, they calculated the vibrational partition function and other relevant thermodynamic properties. It was stated that the modified Mobius square potential can be reduced to the Hua potential model using appropriate potential constant values.

It has been established that the solution for the Schrodinger equation for $l \neq 0$, one has to use a Pekeris-type approximation scheme to deal with the centrifugal term or solve numerically. The following new improved approximation scheme have been used to deal with the centrifugal term (Pekeris-type approximation scheme) [13]:

$$\frac{1}{r^2} = 4\alpha^2 \left( C_0 + \frac{e^{-2\alpha r}}{1-e^{-2\alpha r}} \right) = 4\alpha^2 \left( C_0 + \frac{e^{-2\alpha r}}{1-e^{-2\alpha r}} + \left(\frac{e^{-2\alpha r}}{1-e^{-2\alpha r}}\right)^2 \right) \tag{1c}$$

Greene–Aldrich approximation scheme [11], which is given by

$$\frac{1}{r^2} = \frac{4\alpha^2 e^{-2\alpha r}}{(1-e^{-2\alpha r})^2} \tag{1d}$$

In this article, we are motivated by the numerous studies that have been presented by myriads of researchers in proffering solutions to the SE. However, in the present manuscript we will solve the SE for MMSP with two approximation scheme.

This paper is organized as follows. In Sect. 2, the review of the NU method is presented. In Sect. 3, this method is applied to the radial Schrodinger equation to find the analytical solution with Mobius square potential (MMSP). In Sect. 3, the NU method is also applied to radial Schrodinger equation with Hua potential. In Sect. 4, a brief conclusion is given.

## 2. REVIEW OF NIKIFOROV-UVAROV METHOD

The Nikiforov-Uvarov (NU) method is based on solving the hypergeometric-type second-order differential equations by means of the special orthogonal functions[18]. The main equation which is closely associated with the method is given in the following form [19]

$$\psi''(s) + \frac{\tilde{\tau}(s)}{\sigma(s)}\psi'(s) + \frac{\tilde{\sigma}(s)}{\sigma^2(s)}\psi(s) = 0 \tag{2}$$

Where $\sigma(s)$ and $\tilde{\sigma}(s)$ are polynomials at most second-degree, $\tilde{\tau}(s)$ is a first-degree polynomial and $\psi(s)$ is a function of the hypergeometric-type.

The exact solution of Eq. (2) can be obtained by using the transformation

$$\psi(s) = \phi(s)y(s) \tag{3}$$

This transformation reduces Eq. (2) into a hypergeometric-type equation of the form

$$\sigma(s)y''(s) + \tau(s)y'(s) + \lambda y(s) = 0 \tag{4}$$

The function $\phi(s)$ can be defined as the logarithm derivative

$$\frac{\phi'(s)}{\phi(s)} = \frac{\pi(s)}{\sigma(s)} \tag{5}$$

where $\pi(s) = \frac{1}{2}[\tau(s) - \tilde{\tau}(s)]$ \hfill (5a)

with $\pi(s)$ being at most a first-degree polynomial. The second $\psi(s)$ being $y_n(n)$ in Eq. (3), is the hypergeometric function with its polynomial solution given by Rodrigues relation

$$y^{(n)}(s) = \frac{B_n}{\rho(s)} \frac{d^n}{ds^n}[\sigma^n \rho(s)] \tag{6}$$

Here, $B_n$ is the normalization constant and $\rho(s)$ is the weight function which must satisfy the condition

$$(\sigma(s)\rho(s))' = \sigma(s)\tau(s) \tag{7}$$

$$\tau(s) = \tilde{\tau}(s) + 2\pi(s) \tag{8}$$

It should be noted that the derivative of $\tau(s)$ with respect to $s$ should be negative. The eigenfunctions and eigenvalues can be obtained using the definition of the following function $\pi(s)$ and parameter $\lambda$, respectively:

$$\pi(s) = \frac{\sigma'(s) - \tilde{\tau}(s)}{2} \pm \sqrt{\left(\frac{\sigma'(s) - \tilde{\tau}(s)}{2}\right)^2 - \tilde{\sigma}(s) + k\sigma(s)} \tag{9}$$

where $k = \lambda - \pi'(s)$ \hfill (10)

The value of $k$ can be obtained by setting the discriminant of the square root in Eq. (9) equal to zero. As such, the new eigenvalue equation can be given as

$$\lambda_n = -n\tau'(s) - \frac{n(n-1)}{2}\sigma''(s), n = 0,1,2, \ldots \tag{11}$$

### 3. Bound State Solution with MMSP

The radial Schrodinger equation can be given as:

$$\frac{d^2 R_{nl}}{dr^2} + \frac{2\mu}{\hbar^2}\left[E_{nl} - V(r) - \frac{\hbar^2 \ell(\ell+1)}{2\mu r^2}\right] R_{nl} = 0 \qquad (12)$$

where $\mu$ is the reduced mass, $E_{nl}$ is the energy spectrum, $\hbar$ is the reduced Planck's constant and $n$ and $l$ are the radial and orbital angular momentum quantum numbers respectively (or vibration-rotation quantum number in quantum chemistry). Substituting equation (1a) into equation (12) gives:

$$\frac{d^2 R_{nl}}{dr^2} + \frac{2\mu}{\hbar^2}\left[E_{nl} + \frac{V_0 A^2}{(1-e^{-2\alpha r})^2} + \frac{2V_0 AB e^{-2\alpha r}}{(1-e^{-2\alpha r})^2} + \frac{V_0 B^2 e^{-4\alpha r}}{(1-e^{-2\alpha r})^2} - \frac{\hbar^2 \ell(\ell+1)}{2\mu r^2}\right] R_{nl} = 0 \qquad (13)$$

Substituting Approximation 1(Equation 1d) [24, 25], equation 13 becomes;

$$\frac{d^2 R_{nl}}{dr^2} + \frac{2\mu}{\hbar^2}\left[E_{nl} + \frac{V_0 A^2}{(1-e^{-2\alpha r})^2} + \frac{2V_0 AB e^{-2\alpha r}}{(1-e^{-2\alpha r})^2} + \frac{V_0 B^2 e^{-4\alpha r}}{(1-e^{-2\alpha r})^2} - \frac{2\alpha^2 \hbar^2 \ell(\ell+1)}{\mu}\left(C_0 + \frac{e^{-2\alpha r}}{1-e^{-2\alpha r}} + \left(\frac{e^{-2\alpha r}}{1-e^{-2\alpha r}}\right)^2\right)\right] R_{nl} = 0 \quad (14)$$

$$\frac{d^2 R_{n\ell}(r)}{dr^2} + \frac{2\mu}{\hbar^2 (1-e^{-2\alpha r})^2}\Big[E_{nl}(1 - e^{-2\alpha r})^2 + V_0 A^2 + 2V_0 AB e^{-2\alpha r} + V_0 B^2 e^{-4\alpha r} - \frac{2\alpha^2 \hbar^2 \ell(\ell+1)}{\mu}(C_0(1 - e^{-2\alpha r})^2 + (1 - e^{-2\alpha r}) + e^{-4\alpha r})\Big] R_{n\ell}(r) = 0 \qquad (15)$$

Eq. (15) can be simplified into the form and introducing the following dimensionless abbreviations

$$\begin{cases} -\varepsilon_n = \frac{\mu E_{n\ell}}{2\hbar^2 \alpha^2} \\ \beta = \frac{\mu V_0 B^2}{2\hbar^2 \alpha^2} \\ \gamma = \ell(\ell+1) \\ \delta = \frac{\mu V_0 AB}{\hbar^2 \alpha^2} \\ \eta = \frac{\mu V_0 A^2}{2\hbar^2 \alpha^2} \end{cases} \qquad (16)$$

And using the transformation $s = e^{-2\alpha r}$ so as to enable us apply the NU method as a solution of the hypergeometric type

$$\frac{d^2 R_{n\ell}(r)}{dr^2} = 4\alpha^2 s^2 \frac{d^2 R_{n\ell}(s)}{ds^2} + 4\alpha^2 s \frac{dR_{n\ell}(s)}{ds} \qquad (17)$$

$$\frac{d^2 R_{n\ell}(s)}{ds^2} + \frac{1-s}{s(1-s)}\frac{dR_{n\ell}(s)}{ds} + \frac{1}{s^2(1-s)^2}\Big[-s^2(\varepsilon_n - \beta + \gamma C_0) + s(2\varepsilon_n + \delta + \gamma(2C_0 - 1)) - (\varepsilon_n - \eta + \gamma C_0)\Big] R_{n\ell}(s) = 0 \qquad (18)$$

Comparing Eq. (18) and Eq. (2), we have the following parameters

$$\begin{cases} \tilde{\tau}(s) = 1 - s \\ \sigma(s) = s(1-s) \\ \tilde{\sigma}(s) = -s^2(\varepsilon_n - \beta + \gamma C_0) + s(2\varepsilon_n + \delta + \gamma(2C_0 - 1)) - (\varepsilon_n - \eta + \gamma C_0) \end{cases} \quad (19)$$

Substituting these polynomials into Eq. (9), we get $\pi(s)$ to be

$$\pi(s) = -\frac{s}{2} \pm \sqrt{(a-k)s^2 + (b+k)s + c} \quad (20)$$

where

$$\begin{cases} a = \frac{1}{4} + \varepsilon_n - \beta + \gamma C_0 \\ b = -2\varepsilon_n - \delta - \gamma(2C_0 - 1) \\ c = \varepsilon_n - \eta + \gamma C_0 \end{cases} \quad (21)$$

To find the constant $k$, the discriminant of the expression under the square root of Eq. (20) must be equal to zero. As such, we have that

$$k_\pm = (\delta - \gamma + 2\eta) \pm 2\sqrt{\varepsilon_n - \eta + \gamma C_0}\sqrt{\left(\frac{1}{4} - \delta + \gamma - \eta - \beta\right)} \quad (22)$$

Substituting Eq. (22) into Eq. (20) yields

$$\pi(s) = -\frac{s}{2} \pm \left[\left(\sqrt{\left(\frac{1}{4} - \delta + \gamma - \eta - \beta\right)} + \sqrt{(\varepsilon_n - \eta + \gamma C_0)}\right)s - \sqrt{(\varepsilon_n - \eta + \gamma C_0)}\right] \quad (23)$$

From the knowledge of NU method, we choose the expression $\pi(s)_-$ which the function $\tau(s)$ has a negative derivative. This is given by

$$k_- = (\delta - \gamma + 2\eta) - 2\sqrt{\varepsilon_n - \eta + \gamma C_0}\sqrt{\left(\frac{1}{4} - \delta + \gamma - \eta - \beta\right)} \quad (24)$$

with $\tau(s)$ being obtained as

$$\tau(s) = 1 - 2s - 2\left[\left(\sqrt{\left(\frac{1}{4} - \delta + \gamma - \eta - \beta\right)} + \sqrt{(\varepsilon_n - \eta + \gamma C_0)}\right)s - \sqrt{(\varepsilon_n - \eta + \gamma C_0)}\right] \quad (25)$$

Referring to Eq. (10), we define the constant $\lambda$ as

$$\lambda = (\delta - \gamma + 2\eta) - 2\sqrt{\varepsilon_n - \eta + \gamma C_0}\sqrt{\left(\frac{1}{4} - \delta + \gamma - \eta - \beta\right)} - \frac{1}{2} - \sqrt{\left(\frac{1}{4} - \delta + \gamma - \eta - \beta\right)} + \sqrt{(\varepsilon_n - \eta + \gamma C_0)} \quad (26)$$

Substituting Eq. (26) into Eq. (11) and carrying out simple algebra, where

$$\tau'(s) = -2 - 2\left(\sqrt{\left(\frac{1}{4} - \delta + \gamma - \eta - \beta\right)} + \sqrt{(\varepsilon_n - \eta + \gamma C_0)}\right) < 0 \tag{27}$$

and

$$\sigma''(s) = -2 \tag{28}$$

We have

$$\varepsilon_n = \eta - \gamma C_0 + \frac{1}{4}\left[\frac{\eta - \beta - \left(n + \frac{1}{2} + \sqrt{\frac{1}{4} - \delta + \gamma - \eta - \beta}\right)^2}{\left(n + \frac{1}{2} + \sqrt{\frac{1}{4} - \delta + \gamma - \eta - \beta}\right)}\right]^2 \tag{29}$$

Substituting Eqs. (16) into Eq. (29) yields the energy eigenvalue equation of the modified Mobuis square potential in the form

$$E_{n\ell}^{(Approx.1)} = \frac{2\hbar^2\alpha^2 C_0 \ell(\ell+1)}{\mu} - V_0 A^2 - \frac{\hbar^2\alpha^2}{2\mu}\left[\frac{\frac{\mu V_0 A^2}{2\hbar^2\alpha^2} - \frac{\mu V_0 B^2}{2\hbar^2\alpha^2} - \chi^2}{\chi}\right]^2 \tag{30}$$

The corresponding wave functions can be evaluated by substituting $\pi(s)\_$ $and$ $\sigma(s)$ from Eq. (23) and Eq. (19) respectively into Eq. (5) and solving the first order differential equation. This gives

$$\Phi(s) = s^{\sqrt{\varepsilon_n - \eta + \gamma C_0}}(1-s)^{\frac{1}{2}+\sqrt{\frac{1}{4} - \delta + \gamma - \eta - \beta}} \tag{31}$$

The weight function $\rho(s)$ from Eq. (7) can be obtained as

$$\rho(s) = s^{2\sqrt{\varepsilon_n - \eta + \gamma C_0}}(1-s)^{2\sqrt{\frac{1}{4} - \delta + \gamma - \eta - \beta}} \tag{32}$$

From the Rodrigues relation of Eq. (6), we obtain

$$y_n(s) \equiv N_{n,l} P_n^{\left(2\sqrt{\varepsilon_n - \eta + \gamma C_0},\, 2\sqrt{\frac{1}{4} - \delta + \gamma - \eta - \beta}\right)}(1 - 2s) \tag{33}$$

where $P_n^{(\theta,\vartheta)}$ is the Jacobi Polynomial.

Substituting $\Phi(s)$ $and$ $y_n(s)$ from Eq. (35) and Eq. (37) respectively into Eq. (3), we obtain

$$R_n(s) = N_{n,l} s^{\sqrt{\varepsilon_n - \eta + \gamma C_0}}(1-s)^{\frac{1}{2}+\sqrt{\frac{1}{4} - \delta + \gamma - \eta - \beta}} P_n^{\left(2\sqrt{\varepsilon_n - \eta + \gamma C_0},\, 2\sqrt{\frac{1}{4} - \delta + \gamma - \eta - \beta}\right)}(1 - 2s) \tag{34}$$

Again by using approximation 2 (Equation 1e) and repeating the above procedure, we can consequently obtain the energy eigenvalues as

$$E_{n\ell}^{(Approx.2)} = -V_0 A^2 - \frac{\hbar^2 \alpha^2}{2\mu} \left[ \frac{\frac{\mu V_0 A^2}{2\hbar^2 \alpha^2} - \frac{\mu V_0 B^2}{2\hbar^2 \alpha^2} - \chi^2}{\chi} \right]^2 \quad (35)$$

$$\chi = \left( n + \frac{1}{2} + \sqrt{\frac{1}{4} - \frac{\mu V_0 AB}{\hbar^2 \alpha^2} + \ell(\ell+1) - \frac{\mu V_0 A^2}{2\hbar^2 \alpha^2} - \frac{\mu V_0 B^2}{2\hbar^2 \alpha^2}} \right) \quad (36)$$

4. **CONCLUSION**

In this work, we have investigated non-relativistic problem of Schrodinger equation subject to the Modified Mobuis Square potential respectively within the framework of two approximations scheme. We have obtained exact energy eigenvalues equation and radial wave functions using the NU method. Finally, by choosing appropriate values for potential constants, we have investigated energy eigenvalues for special cases of the Modified Mobuis Square potential.